\documentclass[conference]{IEEEtran}
\IEEEoverridecommandlockouts
\usepackage{cite}
\usepackage{amsmath,amssymb,amsfonts}
\usepackage{csquotes}
\usepackage{algorithmic}
\usepackage{lipsum}
\usepackage{graphicx}
\usepackage{textcomp}
\usepackage{xcolor}
\usepackage{float}
\usepackage{hyperref}

 \makeatletter
\newcommand {\linebreakand}{%
 \end{@IEEEauthorhalign}
  \hfill\mbox{}\par
  \mbox{}\hfill\begin{@IEEEauthorhalign}
}
 \makeatother

\def\BibTeX{{\rm B\kern-.05em{\sc i\kern-.025em b}\kern-.08em
    T\kern-.1667em\lower.7ex\hbox{E}\kern-.125emX}}

\begin{document}

\title{How Many Papers Should You Review? A Research Synthesis of Systematic Literature Reviews in Software Engineering}

\author{\IEEEauthorblockN{Xiaofeng Wang}
\IEEEauthorblockA{\textit{Free University of Bozen-Bolzano}\\
Bolzano, Italy \\
xiaofeng.wang@unibz.it}
\and
\IEEEauthorblockN{Henry Edison}
\IEEEauthorblockA{\textit{Blekinge Institute of Technology}\\
Karlskrona, Sweden \\
henry.edison@bth.se}
\linebreakand 
\IEEEauthorblockN{Dron Khanna}
\IEEEauthorblockA{\textit{Free University of Bozen-Bolzano}\\
Bolzano, Italy \\
dron.khanna@unibz.it}
\and
\IEEEauthorblockN{Usman Rafiq}
\IEEEauthorblockA{\textit{Free University of Bozen-Bolzano}\\
Bolzano, Italy \\
urafiq@unibz.it}
}

\IEEEpubid{
	\begin{minipage}{\textwidth}\ \\[12pt]
    978-1-6654-5223-6/23/\$31.00 \copyright 2023 IEEE
	\end{minipage}
} 

\maketitle
\begin{abstract}
[Context] Systematic Literature Review (SLR) has been a major type of study published in Software Engineering (SE) venues for about two decades. However, there is a lack of understanding of whether an SLR is really needed in comparison to a more conventional literature review. Very often, SE researchers embark on an SLR with such doubts. We aspire to provide more understanding of when an SLR in SE should be conducted. [Objective] The first step of our investigation was focused on the dataset, i.e., the reviewed papers, in an SLR, which indicates the development of a research topic or area. The objective of this step is to provide a better understanding of the characteristics of the datasets of SLRs in SE. [Method] A research synthesis was conducted on a sample of 170 SLRs published in top-tier SE journals. We extracted and analysed the quantitative attributes of the datasets of these SLRs. [Results] The findings show that the median size of the datasets in our sample is 57 reviewed papers, and the median review period covered is 14 years. The number of reviewed papers and review period have a very weak and non-significant positive correlation. [Conclusions] The results of our study can be used by SE researchers as an indicator or benchmark to understand whether an SLR is conducted at a good time.
\end{abstract}

\begin{IEEEkeywords}
SLR, Systematic Literature Review, Methodological Study, Research Synthesis, Software Engineering
\end{IEEEkeywords}

\section{Introduction}
\label{sec:intro}
Systematic literature reviews (SLRs) have a strong presence in Software Engineering (SE) literature, and the number of SLR studies has grown steadily in the last two decades \cite{ali18}. SLRs, like any research, should be performed carefully, following rigorous processes, and results should be reported and interpreted appropriately. They require considerably more effort than traditional literature reviews \cite{kitchenham2007guidelines}. Therefore, SE researchers should not commit to conducting an SLR without understanding whether it is worth doing.

The worthiness can be understood from different perspectives, among which an important one is timing, i.e., when is the appropriate time to conduct an SLR? Or is there an appropriate time at all? Despite several guidelines and tertiary studies on SLRs in SE \cite{kitchenham2007guidelines, kitchenham2009systematic, kitchenham2010systematic}, no clear indications are provided on the right time to conduct an SLR on a research question, area, or phenomenon in the SE research field. We aspire to fill this observed knowledge gap. 

As the first step of our research, we investigated the datasets, i.e., the reviewed papers, in SLRs in SE. We assumed that analysing the dataset of an SLR can reveal the development status of a research topic or area when an SLR was conducted. Therefore, we asked the following research question: 

\begin{displayquote} \textbf{\textit{What are the characteristics of the datasets of SLRs in SE?}}
\end{displayquote}

To answer the research question, we conducted a research synthesis on a sample of SLRs published in top-tier SE journals. For each of the SLR studies in the sample, we extracted relevant data on the reviewed papers, including the number of reviewed papers and the period covered by these reviewed papers. The collected data was analysed through multiple angles to reach the answer to the posed research question. The findings reported in the paper provide insights into the characteristics of the datasets used by SLRs in SE. SE researchers can take our findings as an indicator or benchmark to understand whether an SLR is conducted at a good time. 

The rest of the paper is organised as follows. Section \ref{sec:lit_review} provides a review of the guidelines and tertiary studies that are relevant to our study. The data collection process we followed to build our study sample is described in Section \ref{sec:research_method}. The following section, Section \ref{sec:results}, reports the findings, which are discussed in Section \ref{sec:discussion}. Lastly, in Section \ref{sec:next_steps}, we outline the next steps of our research on understanding the temporal aspects of SLRs in SE.

\section{Related Work}
\label{sec:lit_review}
The widely used guidelines of SLRs in SE are provided in \cite{kitchenham2007guidelines}, in which the reasons for performing SLRs and their importance are argued. Later on, guidelines for the search strategy to update SLRs in SE are provided in \cite{wohlin2020guidelines}. Recently, Kitchenham et al. \cite{kitchenham2022segress} presented an integrated set of guidelines to address reporting problems in secondary SE studies. Apart from these guidelines, several tertiary studies in SE exist in the literature. These studies assess the impact of SLRs and provide an annotated catalogue of SLRs (e.g., \cite{kitchenham2009systematic, kitchenham2010systematic}), record the reported experiences of conducting SLRs for the benefit of new researchers \cite{Imtiaz2013}, or review SLRs in a specific SE area (e.g., Software Engineering Education \cite{budgen2020support}).

Few existing guidelines or tertiary studies in SE suggest the appropriate time to conduct an SLR on a research question or topic. The study of Mendes et al. \cite{mendes2020update} is the only one that we are aware of investigating the timing aspect of SLRs in SE. Their goal is to understand when is the appropriate time to update SLRs in SE. Using a decision framework employed in other fields, they analysed 20 SLRs which are updates of previously conducted SLRs. The study finds that 14 of the 20 updated SLRs need not be conducted. 

The work of Mendes et al. \cite{mendes2020update} provides a good motivation to examine the necessity of conducting first-time SLRs in SE, which is not investigated by these authors or in any existing SE literature as far as we are aware of. More specifically to the focus of this paper, no suggestion is provided on how many papers should be reviewed in an SLR in SE. Understandably, suggestions like this are difficult to offer since each research topic or area has a different development pace, has a different number of researchers working on it, and therefore accumulates evidence and knowledge at a different speed. Nevertheless, it would be useful to have an overall understanding of the datasets used by SLRs in SE, since a dataset, i.e., reviewed papers, in an SLR represents the knowledge accumulated on the research topic under the investigation.

\section{Research Approach}
\label{sec:research_method}
To answer the research question, we employed  research synthesis. Research synthesis is an umbrella term referring to methods used to summarise, integrate, combine, and compare the findings of different studies on a particular topic or research question \cite{woods05,cooper09,cruzes11}. Research synthesis aims at analysing and evaluating multiple studies to integrate and provide new interpretative explanations about them \cite{cruzes11}. We conducted a research synthesis of a sample of SLRs in SE, focusing on the datasets used in these SLRs to investigate how many papers should be reviewed in an SLR.

\subsection{Data collection}
\subsubsection{Search strategy}
Even though we were not conducting an SLR study, we followed the search strategy defined in \cite{kitchenham2007guidelines} to build our sample. We did not attempt to search for all relevant SLRs in SE exhaustively but rather to sample enough studies for analysis. Therefore, we focused on SLRs published in top-tier SE journals as identified by Wong et al. \cite{wong21}. This is a trade-off between considering as much literature as possible and at the same time accumulating and extracting reliable information. As reported in \cite{ali18}, more than 600 SLRs were published between 2004 and 2016, and there is a trend that the number has been growing since. Therefore, the number of SLRs published in journals can already provide enough data for the first step of our study.

To build our search string, we combined the journals' titles with the synonyms of ``systematic literature reviews'' \cite{biolchini07}. Our generic search string is:
\begin{displayquote}
\textit{(``systematic review" OR ``research review" OR ``research synthesis" OR ``research integration" OR ``systematic overview" OR ``systematic research synthesis" OR ``integrative research review" OR ``integrative review" OR ``systematic literature review" OR ``literature review") AND (``Information and Software Technology" OR ``Journal of Systems and Software" OR ``IEEE Software" OR ``IEEE Transactions on Software Engineering" OR ``Software: Practice and Experience" OR ``Software Testing, Verification and Reliability" OR ``Transactions on Programming Languages and Systems" OR ``Transactions on Software Engineering and Methodology" OR ``Journal of Software: Evolution and Process" OR ``International Journal on Software Tools for Technology Transfer" OR ``Empirical Software Engineering")}
\end{displayquote}

We ran the search string in Scopus on Feb 24, 2023, and retrieved 412 published papers. Each paper was inspected by two authors to decide whether it is an SLR study or follows SLR guidelines. In the cases where the two authors did not agree on the decision, a third author's vote was required. We excluded the studies that do not follow SLR guidelines (e.g., conventional/ad-hoc reviews). We also excluded mapping studies, grey literature reviews, multi-vocal literature reviews, tertiary studies or SLR updates. Some studies published in IEEE Software are typically summaries of existing SLRs that have already been published in other venues. We checked the venues where the original SLRs were published. We included the original SLRs in our sample if the venues are among the journals we used to search for SLRs. 

\subsubsection{Data extraction}
A key element of an SLR is dataset, i.e., the papers reviewed in an SLR.  There are various facets of a dataset that could be relevant to our study. In this first step, we focused on the following three facets:

\begin{itemize}
    \item The number of reviewed papers in the SLR;
    \item The earliest publication year of the reviewed papers; and
    \item The latest publication year of the reviewed papers.
\end{itemize}

If an SLR does not report the information above or provide detailed information on how to get them, we excluded it from our sample. The unit of analysis in our study is the SLR study itself. Therefore, if two SLR studies were conducted and reported in one paper, we considered two data points from that paper. Moreover, if a published paper contains both an SLR and other review studies (e.g., systematic mapping study) or empirical studies (e.g., case studies, experiments, etc.), we only included the paper if we were able to extract the SLR-related data. 

For each of the identified SLR studies, the meta-data of the paper in which it is published (e.g., title, publication year, authors, publication venue) were extracted automatically through the ``export'' feature of Scopus. 

Ultimately, we collected data from 170 SLRs, constituting our final data analysis sample. The final version of the dataset is accessible through a publicly available repository \cite{Wang2023}.

\subsection{Data analysis}
In the data analysis phase, apart from the meta-data of the publications containing the SLRs, we defined the following two variables directly related to the dataset of an SLR:
\begin{itemize}
    \item NoRP: Number of Reviewed Papers in an SLR; and
    \item RPC: Review Period Covered by the reviewed papers. \textit{RPC = the latest publication year of the reviewed papers in an SLR - the earliest publication year of the reviewed papers in an SLR + 1}
\end{itemize}

After obtaining the descriptive statistics (min, max, median, mean and standard deviation) of NoRP and RPC, we explored whether there was any relation between the two variables. That is, to understand whether the number of reviewed papers in an SLR can be indicated by how long the research topic under the study has been explored. 

\section{Results}
\label{sec:results}
\subsection{Sample overview}
\begin{figure}[h]
\centering
    \includegraphics[width=0.5\textwidth, height=5cm]{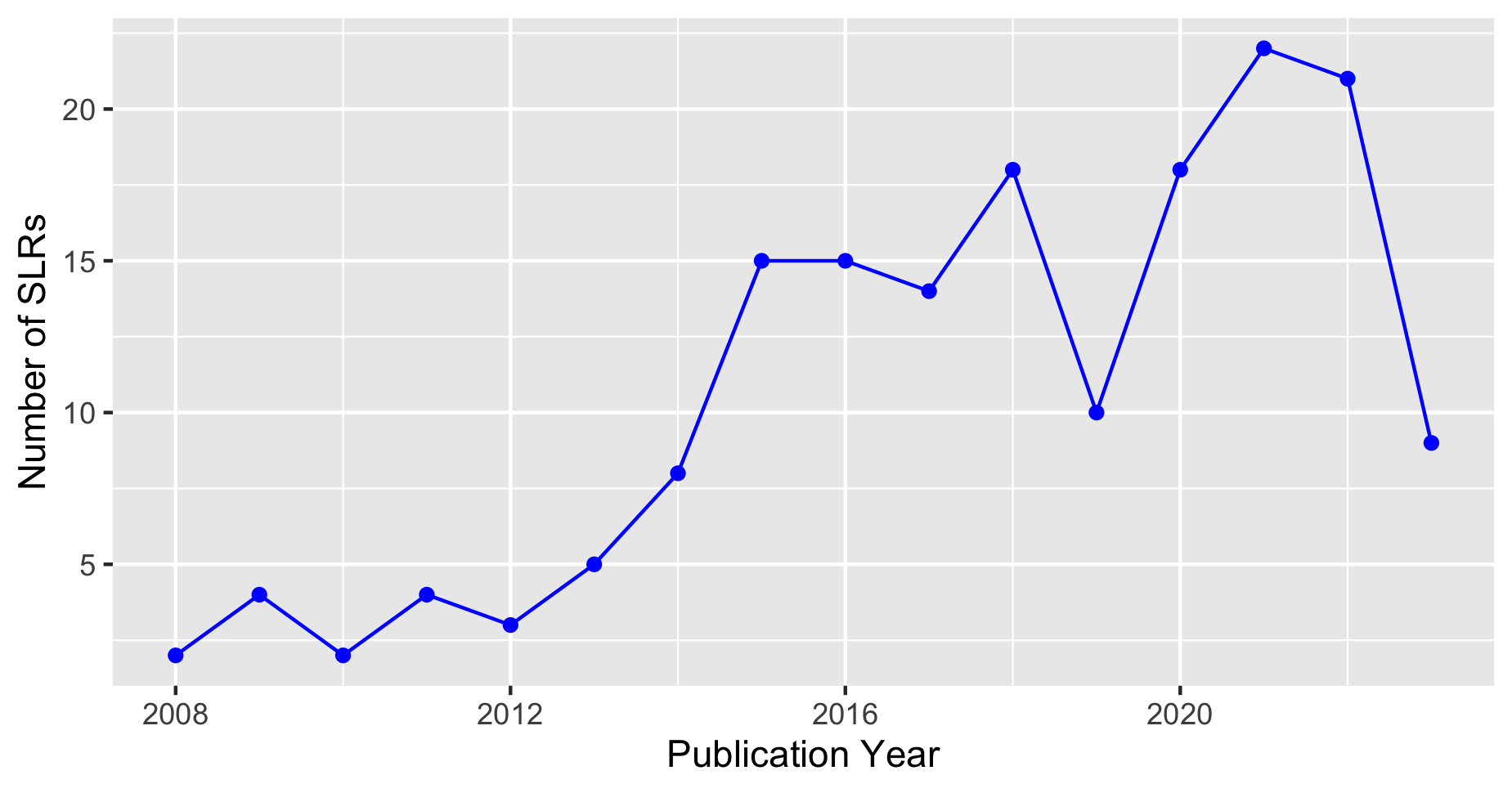}
    \caption{Number of SLRs per year (n = 170)}
    \label{fig:SLRs_per_year}
\end{figure}

Before reporting the results related to the two variables, we provide an overview of the 170 SLRs in our sample, as shown in Fig. \ref{fig:SLRs_per_year} and Table \ref{tab:pub_venue}. 

Fig. \ref{fig:SLRs_per_year} shows the distribution of the SLRs across the years. In our sample, the two earliest SLRs were published in IST in 2008. The number of SLRs published in top-tier journals has been growing over the years, despite having small dips in certain years.

Table \ref{tab:pub_venue} shows the distribution of these SLRs across the journals. It can be seen from Table \ref{tab:pub_venue} that the \textit{Journal of Systems and Software} has the most SLRs (70), followed by \textit{Information and Software Technology} (40). \textit{Journal of Software: Evolution and Process} and \textit{Empirical Software Engineering} have similar numbers of SLRs (18 and 16, respectively).
\begin{table}[h]
    \centering
    \caption{The distribution of the SLRs across the journals}
    \label{tab:pub_venue}
    \begin{tabular}{p{6.4cm}p{1.5cm}}
    \hline
    Journal name & No. of SLRs \\
    \hline
        Journal of Systems and Software (JSS) & 70 \\
        Information and Software Technology (IST) & 40 \\
        Journal of Software: Evolution and Process & 18 \\
        Empirical Software Engineering & 16 \\
        IEEE Transactions on Software Engineering (TSE) & 12 \\
        Software - Practice and Experience & 6 \\
        ACM Transactions on Software Engineering and Methodology (TOSEM) & 3\\
        Software Testing Verification and Reliability & 3 \\
        IEEE Software & 1 \\ 
        International Journal on Software Tools for Technology Transfer & 1 \\
\hline
Total & 170\\
         \hline
    \end{tabular}
\end{table}

\subsection{Characteristics of the datasets of SE SLRs}
Table \ref{tab:description_var} lists the descriptive statistics of the two variables, NoRP and RPC. 

\begin{table}[h]
    \centering
    \caption{The descriptive statistics of the two variables}
    \label{tab:description_var}
    \begin{tabular}{p{3cm}p{1cm}p{1cm}p{1cm}}
    \hline
         &NoRP (n=170) &NoRP (n=166)&RPC (n=170)  \\
         \hline
         Min&6&6&2\\
         Max&925&250&41\\
         Median&57&56.5&14\\
         Mean&80.59&69.29&15.34\\
         Standard deviation (sd)&95.60&50.24&8.22\\
         \hline
    \end{tabular}
\end{table}

\begin{figure*}[h]
    \centering
    \includegraphics[scale=0.16]{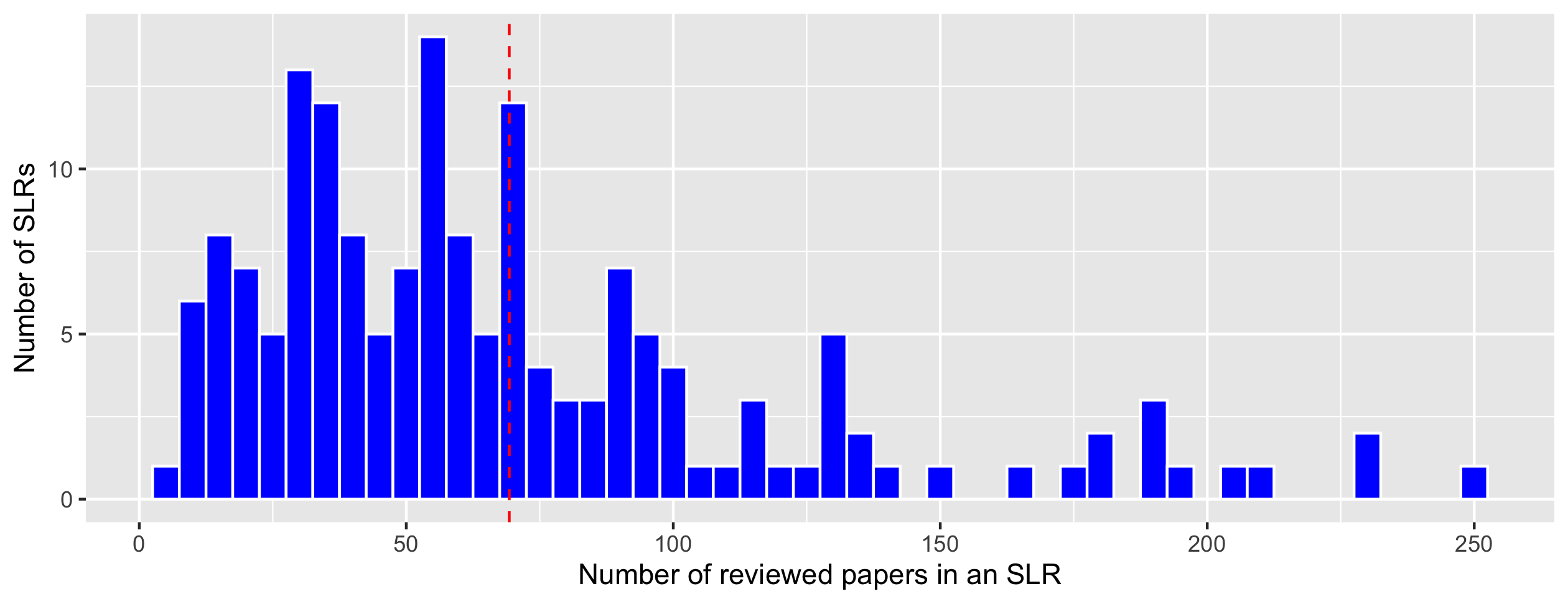}
    \caption{The distribution of the number of reviewed studies in SLRs (n=166)}
    \label{fig:NoRP_histogram}
\end{figure*}

As shown in the first column ``NoRP (n=170)'' of Table \ref{tab:description_var}, the number of reviewed papers, or the size of the datasets of the SLRs, varies greatly (sd=95.60). The minimum number of reviewed papers is 6 (in one SLR), and the maximum is 925 (in one SLR). The median size is 57, and the mean value is 80.59, which means the number of reviewed papers is right-skewed. Indeed, after removing the outliers (the four largest numbers of NoRP) to make Fig. \ref{fig:NoRP_histogram} more readable (otherwise, the majority of the data points will be squeezed into a small area of the diagram), the difference between the median and mean values is reduced, as well as the standard deviation, as shown in the second column ``NoRP (n=166)'' of Table \ref{tab:description_var}.

\begin{figure*}[h]
    \centering
    \includegraphics[scale=0.15]{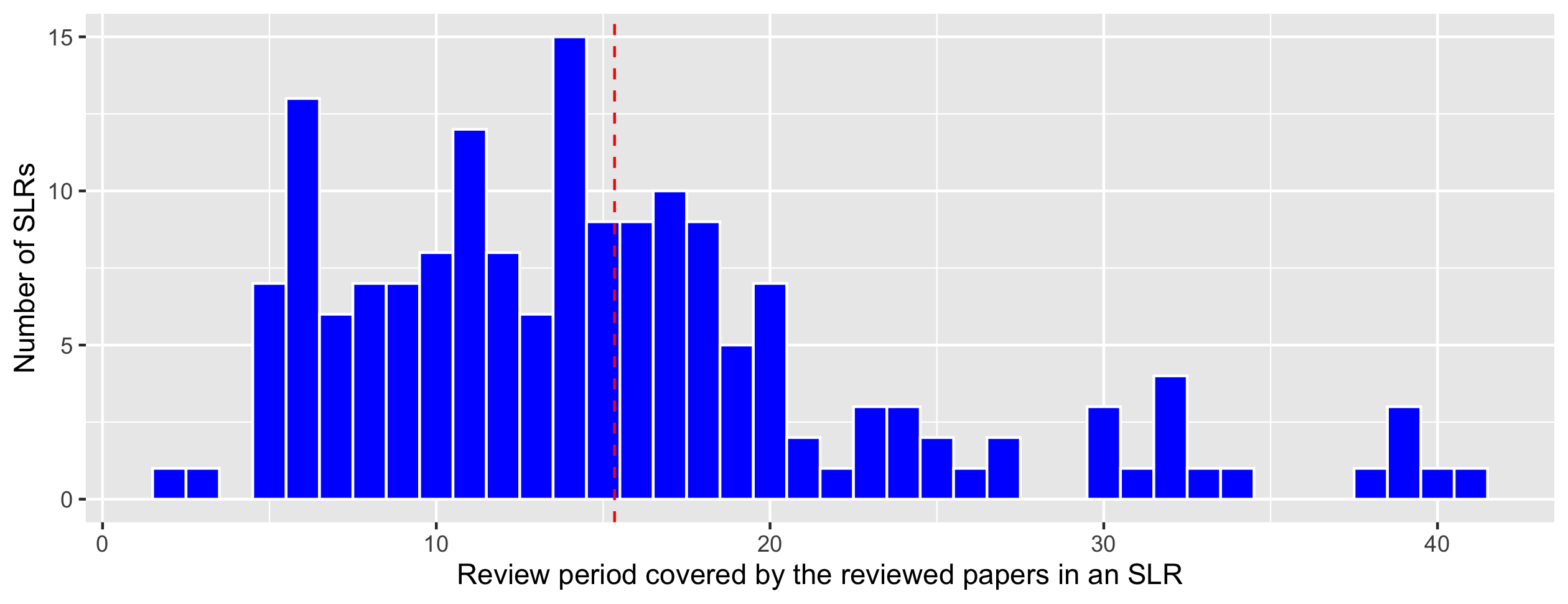}
    \caption{The distribution of the year span of reviewed studies in SLRs (n=170)}
    \label{fig:RPC_histogram}
\end{figure*}

To show the distribution of NoRP more clearly, we plotted the histogram using the sample of 166 SLRs, as shown in Fig. \ref{fig:NoRP_histogram}. The red line indicates the mean value. It can be seen in Fig. \ref{fig:NoRP_histogram} that the dataset sizes ranging from 53 to 57 reviewed papers are most common, used by fourteen SLRs. The other common size ranges are between 28 and 32 (thirteen SLRs), between 33 and 37 (twelve SLRs), and between 68 and 72 reviewed papers (twelve SLRs). 

As shown in the third column, ``RPC (n=170)'' of Table \ref{tab:description_var}, this variable's median and mean values converge to 14 years, with a standard deviation of 8.22. The longest review period covered by the reviewed papers in an SLR is 41 years. The SLR with the longest review period was published in TSE in 2021. One hundred and sixty-six papers were reviewed in this SLR, ranging from 1977 to 2017. What is somehow surprising is the shortest review period (min value of RPC), which is 2 years. The SLR with the shortest review period was published in \textit{Software Testing Verification and Reliability} in 2014. Despite the short review period, the number of reviewed papers is fifty-four, close to the median dataset size. These fifty-four reviewed papers were published between 2009 and 2010. 

Fig. \ref{fig:RPC_histogram} shows the distribution of RPC, the review periods covered by the reviewed papers in SLRs, using the sample of 170 SLRs. No outlier is perceived since all values are within a reasonable range (between two and 41). The red line indicates the mean value. As shown in Fig. \ref{fig:RPC_histogram}, fifteen SLRs have reviewed the papers published within 14 years, which is the most common review period covered and also the median value of RPC. The next common review period covered is 6 years (thirteen SLRs have this review period), followed by 11 years (twelve SLRs).

\begin{figure*}[h]
\centering
    \includegraphics[scale=0.15]{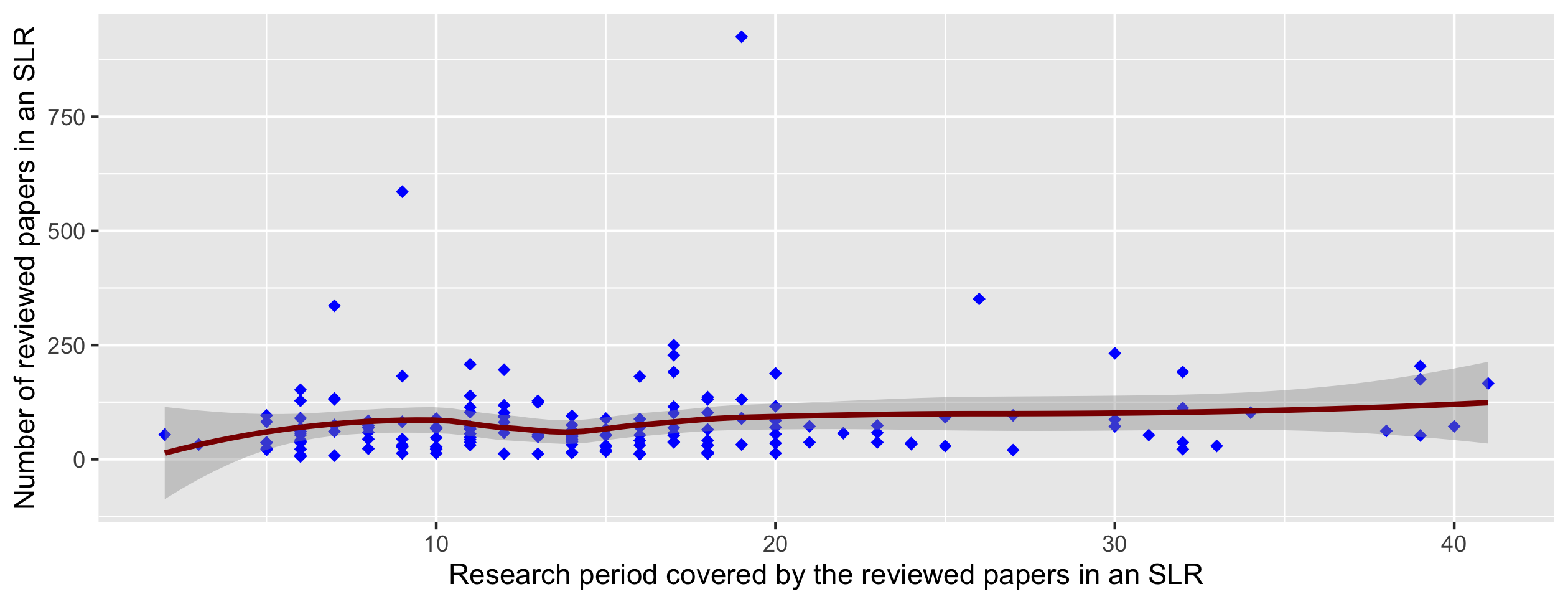}
    \caption{The relation between the number of reviewed papers and review period covered in SLRs (n = 170)}
    \label{fig:NoRPvsRPC}
\end{figure*}

\begin{figure*}[h]
    \centering
    \includegraphics[scale=0.15]{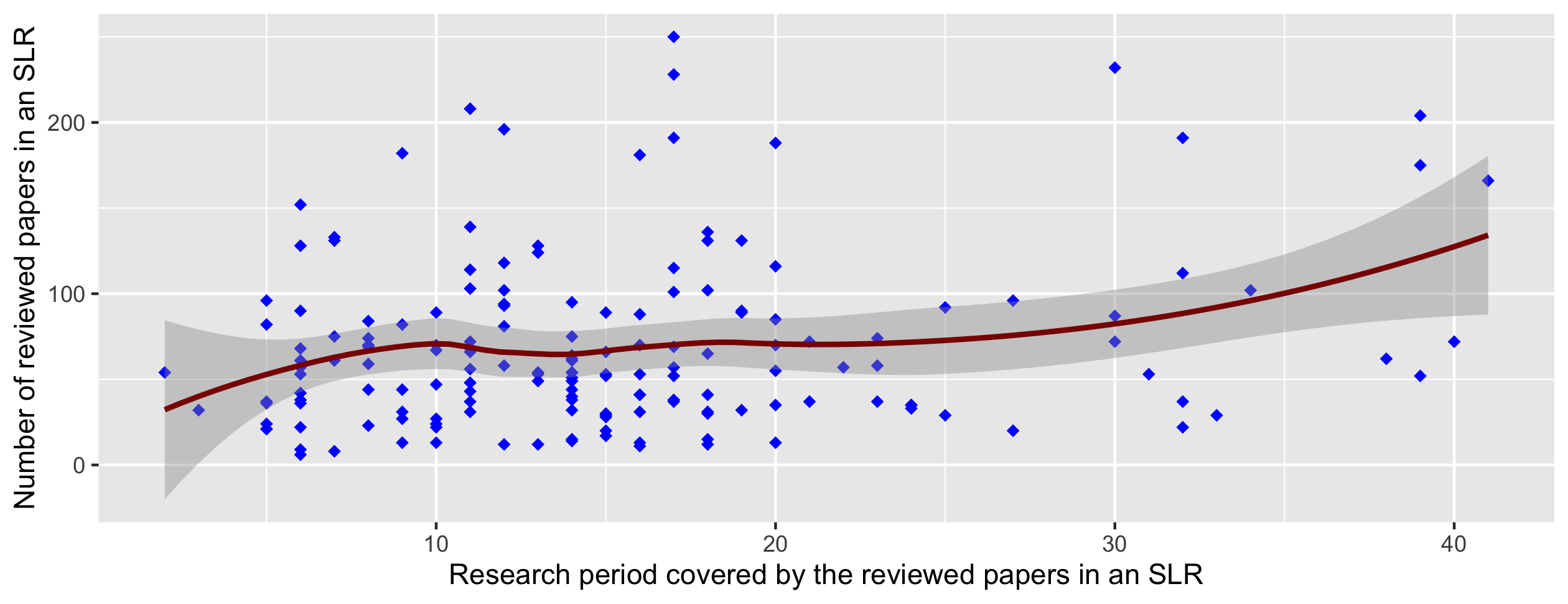}
    \caption{The relation between the number of reviewed papers and review period covered in SLRs (n = 166)}
    \label{fig:NoRPvsRPC2}
\end{figure*}

Fig. \ref{fig:NoRPvsRPC} is the scatterplot of the two variables (NoRP vs RPC) using the sample of 170 SLRs. It shows no observable relationship between the number of reviewed papers in an SLR and the review period covered by that collection of reviewed papers. The scatterplot can be better observed using the sample of 166 SLRs as shown in Fig. \ref{fig:NoRPvsRPC2}. Using both samples, we tested the correlation between NoRP and RPC. Since the two variables are not normally distributed (based on the results of the Sharpiro-Wilk test\cite{hanusz2016shapiro}), we tested their correlation using the Spearman rank correlation coefficient\cite{zar2014spearman} with a 0.95 confidence level. For the sample of 170 SLRs, the results indicate a very weak positive correlation between the two variables (rho = 0.1310, p-value = 0.0886). Similar results were obtained using the sample of 166 SLRs (rho = 0.1357, p-value = 0.0814). However, in both cases, the p-value is above 0.05, which indicates that there is no sufficient evidence to support the correlation between the two variables in both samples.

\section{Discussion}
\label{sec:discussion}
The quantitative analysis conducted on the datasets used by the SLRs in our sample shows that there is no single magic number that SE researchers can rely on to decide whether it is an appropriate time to conduct an SLR. It depends evidently on the research question or topic under investigation. However, the median number of reviewed papers in the SLRs (57) and the typical review period covered (14 years) can serve as a first useful indicator or benchmark to evaluate whether the research on a given topic has accumulated enough studies that warrant an SLR. SE researchers can estimate the dataset they will obtain or compare what they have already obtained to understand whether they are dealing with a smaller or larger dataset than the average ones used by the SLRs in SE. They should be more cautious when the dataset is extremely small or large, which may signal a potential issue in the literature search or inclusion/exclusion processes. Additionally, when the number is extremely small, it may mean that the research field is not mature enough, and an SLR is not needed at that point in time. On the contrary, when the number is extremely large, it indicates that the SLR should have been conducted earlier.

One major limitation of our study is that we constrained our SLR sampling to those published in a selected list of top-tier SE journals. We did not include SLRs published in SE conferences. Therefore, the findings cannot be generated for the SLRs published in those venues.  Another limitation is that we used the Number of Reviewed Papers (NoRP) as an indicator. This number is only obtainable after the relevant papers are retrieved, and inclusion/exclusion criteria are applied, which means a significant amount of effort has already been invested before the NoRP can be known. This limits the usefulness of NoRP as an early-stage indicator of ``when'' to conduct an SLR.

\section{Next Steps and Future Work}
\label{sec:next_steps}
This paper reports the initial findings of our study on the temporal aspects of SLRs in SE. Our eventual goal is to understand when it is an appropriate time to conduct an SLR on an SE research topic. In the first step, we used the number of reviewed papers and review period covered by these papers as the indicators. In the next steps, we will investigate other data, e.g., the number of retrieved papers after applying the search string (assuming a good one), as an earlier indicator on whether an SLR is conducted in a timely manner.

We also need to explore the factors that affect the size of SLR datasets, such as the number of libraries used in the search and the search strategies used (such as closed vs. open period). Additionally, we will collect more data about different facets of the dataset of an SLR, the distribution of the reviewed papers over years and venues, and the types of papers included in a dataset (conference or journal paper, research methodology used, and so on). We will explore the patterns in these data and relations among different facets. 

Another venue for future work is to broaden our sample by collecting and analysing the SLRs published in SE conferences. By contrasting and comparing the SLRs published in these two different types of venues, we can improve the generalisability of our findings.

Our study focused only on the quantitative SLR data. In the future, qualitative analysis can be conducted on SLRs. For example, one can investigates which SE topics have been systematically reviewed and published. One can also map the topics of SLRs to the SE knowledge areas \cite{bourque14} to provide a bigger picture of SE research and its change over time. This could help SE researchers to find the relevant SLRs on their topics and decide if an SLR on their topic is needed. 

Even though we focused on SLR, we believe our research question is relevant to other literature review methods, such as systematic mapping studies or multivocal reviews. Therefore, researchers could replicate our approach to advance our knowledge in these related areas.

\section*{Acknowledgement}
This work has been supported by ELLIIT; the Swedish Strategic Research Area in IT and Mobile Communications.

\bibliographystyle{IEEEtran}
\bibliography{IEEEabrv,References}

\end{document}